# Surface Dynamic Deformation Estimates from Local Seismicity: the Itoiz Reservoir, Spain.

Miguel A. Santoyo[1], Patricia Martínez-Garzón[2], Antonio García-Jerez[3,4] and Francisco Luzón [3,5].

1. Instituto de Geofísica-Unidad Michoacán. Campus Morelia, Universidad Nacional Autónoma de México. México. santoyo@geofisica.unam.mx. Tel: (52) 55-56237862, Fax: (52) 4433222777

2. Helmholtz-Centre Potsdam. GFZ German Research Centre for Geosciences. Potsdam, Germany.

3. Instituto Andaluz de Geofísica, IAGPDS. Universidad de Granada. Granada, Spain.

4. Departamento de Física Teórica y del Cosmos. Universidad de Granada. Granada. Spain.

5. Grupo de Geofísica Aplicada. Departamento de Química y Física. Universidad de Almería. Almería, Spain.

## ABSTRACT

We analyze the ground motion time histories due to the local seismicity near the Itoiz reservoir to estimate the near-source, surface 3D displacement-gradients and dynamic deformations. The seismic data were obtained by a semi-permanent broadband and accelerometric network located on surface and at underground sites. The dynamic deformation field was calculated by two different methodologies. First, by the Seismo-Geodetic method using the data from a three-station micro-array located close to the dam; second, by Single-Station estimates of the displacement gradients. The dynamic deformations obtained from both methods were compared and analyzed in the context of the local free-field effects. The shallow 1D velocity structure was estimated from the seismic data by modeling the body wave travel times. Time histories obtained from both methods results quite similar in the time-window of body-wave arrivals. The strain misfits between methods vary from 1.4% to 35.0% and rotational misfits vary from 2.5% to 36.0%. Amplitudes of displacement gradients vary in the range of $10^{-8}$ to $10^{-7}$ strains. From these results, a new scaling analysis by numerical modeling is proposed in order to estimate the peak dynamic deformations for different magnitudes, up to the expected maximum $M_W$ in the region (M5.5). Peak dynamic deformations due to local $M_W5.5$ earthquakes would reach amplitudes of $10^{-5}$ strain and $10^{-3}$ radians at the Itoiz dam. The Single-Station method shows to be an adequate option for the analysis of local seismicity, where few three-component stations are available. The results obtained here could help to extend the applicability of these methodologies to other sites of engineering interest.

**KEY WORDS**: Dynamic deformations from earthquakes. Peak dynamic strains and rotations. Seismo-Geodetic method. Single Station estimates. Itoiz reservoir.

# 1 INTRODUCTION

Damaging effects produced by earthquakes are the result of peak ground accelerations, large dynamic deformations (strains and rotations) and other strong ground motion variations. While most of the damage may be explained by high ground accelerations, dynamic deformations frequently account for much of the ground failures and life-line damages due to earthquakes (e.g. Ariman and Hamada 1981; Singh et al. 1997). One example of this were the extensive damages produced by the January 17, 1994 Northridge, California earthquake, where a moderate-sized rupture ($M_w6.7$) generated strong non-linear soil response at many locations in the San Fernando Valley (e.g. Todorovska and Trifunac 1996; Schiff 1997). Another significant example was the september 19, 1985 Michoacan earthquake ($M_w8.1$), causing unprecedented damage to Mexico City water-supply system (see, e.g., Rosenblueth and Meli 1986; Rosenblueth and Ovando 1991). Much of these damages were attributed to the large surface dynamic deformations observed in the lake-bed zone during the coseismic strong motion (e.g. Singh et al. 2007)

Dynamic deformations can be also directly related with the damage in large engineering structures as dams or bridges due to earthquakes (e.g. Gazetas 1987; Çelebi et al. 1999; Papalou and Bielak 2004; Wang et al. 2013). Two significant examples of this were the failures produced in the Hsinfengkiang dam in China and the Koyna dam in India, both of them being concrete gravity-dams. In the first case, an $M_s6.1$ earthquake occurred on March 19, 1962, two years after the beginning of impoundment, approximately 5.0 km from the Hsinfengkiang dam. This earthquake produced large cracks on an overhead section of two different structural non-overflow blocks, where dynamic deformations may have played a relevant role (Bolt and Cloud 1974). In the Koyna dam case, an $M_s6.4$ earthquake occurred on December 11, 1967, approximately 15.0 km from the dam. There, the ground motions produced several cracks in both faces of various non-overflow blocks of the structure. These damages were associated not only with the ground accelerations but also with the strong coseismic deformations (e.g. Chopra and Chakrabarti 1973; Paul et al. 1996). An extreme example of strong damages in a concrete gravity dam due to earthquake deformations was the Shih-Kang dam, which was severely shaken during the September 21, 1999, Chi-Chi, Taiwan earthquake (Mw7.6). In this case, the dam structure was directly intersected by the fault rupture. This earthquake produced in some places close to the dam, maximum ground-surface permanent vertical displacements of 11.0 m (e.g. Lee et al. 2001). Damages in the dam structure were extensive, reaching differential vertical displacements in the wall structure of 7.8 m (e.g. Shin and Teng 2001). Although strong damages in dams due to earthquakes are relatively rare, there is a considerable interest on the evaluation of dynamic deformations near dams and other large engineering structures due to the mutual relationship between subsoil strains and stresses.

Free-field dynamic deformations induced by earthquakes have been studied in the past for different regions and events (e.g. Spudich et al. 1995; Bodin et al. 1997; Gomberg 1997; Gomberg and Felzer 2008; Santoyo 2014), as well as their effects on the dynamic triggering of earthquakes in the source regions (e.g. Gomberg and Johnson 2005; Hill and Prejean 2013). However, despite of their relevance for seismic engineering, the analysis of free-field dynamic deformations from seismicity near large infrastructures has not been extensively studied.

Our analysis is focused on the computation of the free-field dynamic deformations near the Itoiz reservoir, which is a newly constructed concrete gravity-dam, located in Navarra, northern Spain (Figure 1). This dam stores the water from the Irati and Urrobi rivers, and it has a total height of 121 m from the streambed, a total length of 525 m and a maximum storage capacity of $4.2x10^8\,m^3$. We calculate and compare in this work the seismic dynamic deformation field by means of two different methods. First, the Seismo-Geodetic method (Bodin et al. 1997) which uses the seismic data from a micro-array of seismic stations; second, by Single Station estimates of the displacement gradients (e.g. Gomberg 1997) assuming the incidence of body waves (S wavefield) propagating through the seismic recording site. Based of these results, a scaling analysis by a numerical modeling of synthetic displacement gradients fitting observations, is performed in order to estimate the peak dynamic deformations for different magnitudes, up to the expected maximum $M_W$ observed in the region (M5.5).

## 1.1 Tectonic and seismic setting

The Itoiz reservoir is located in the Western Pyrenees, which is a geological structure with a roughly E–W orientation belonging to the Pyrenean Range (Figure 1). This region is divided in the North Pyrenean Zone (NPZ), the Paleozoic Axial Zone (PAZ) and the South Pyrenean Zone (SPZ). The NPZ is a Mesozoic unit (156-85Ma) separated from the PAZ by the North Pyrenean Fault (NPF), which is a transform fault system characterized by multiple fault segments, and running in an E–W direction along the entire Pyrenean range (e.g. Larrasoaña et al. 2003). The trace of the NPF system runs towards the western edge of the Pyrenean Belt, up to the Paleozoic Basque Massifs (PBM; Choukroune 1992). The SPZ constitutes the external part of the Pyrenean Belt which is essentially a Tertiary unit (85-25Ma) with a mainly E–W trend of fault systems. Since the Quaternary (2.5Ma) to present time, the active faults in the western side of the Pyrenean range appear to behave mainly with strike slip to normal faulting displacements (e.g. Herraiz et al. 2000; Chevrot et al. 2011). Most of these show **P** axes with a NW–SE general orientation (Ruiz et al. 2006a). The recent stress field in this region is not homogeneous at small and local scales. Focal mechanisms in the western part of the Pyrenean range, suggest trends varying from NW–SE to NNW–SSE (e.g. Ruiz et al. 2006b; Santoyo et al, 2010), with mainly extensional focal mechanisms.

In the western part of the Pyrenees, the seismicity during the second half of the last century appears relatively sparse with small to moderate magnitudes (up to M=5.5) and shallow depths (e.g. Rivas-Medina et al. 2012; Souriau et al. 2014). However, at least six destructive earthquakes have been reported in this area (e.g. Martínez-Solares and Mezcua 2003; Benito et al. 2008). Rivas-Medina et al. (2012) obtained from PSHA analysis, peak ground accelerations for the Itoiz dam site of PGA=0.10g for 975 years of return period (RP) and PGA=0.17g for 4,975 years RP. They also estimated maximum spectral accelerations SA(0.1s)=0.27g for 975 years RP and SA(0.1s)=0.44g for 4,975 years RP. Few clustered seismic series in the region had been reported before the dam construction; however, eight months after the beginning of impoundment, a clustered seismic series began on September 2004 in the same zone. Different works have studied the relation of the water level change of the reservoir lake with this seismicity (Ruiz et al. 2006b; Durá-Gómez and Talwani 2010; Luzón et al. 2010; Santoyo et al. 2010). The series was headed by an Mw=4.7 mainshock and followed by at least 350 moderate and small aftershocks (e.g. Jiménez et al. 2009; Santoyo et al. 2010; Jiménez and Luzón, 2011). The mainshock and the largest aftershock were widely felt in this region and western Pyrenees. Ruiz et al. (2006a) reported maximum intensities of EMS IV in Pamplona City and intensity EMS V in the Itoiz dam; nonetheless, no significant damages were reported during these events. The University of Almería installed in 2008 a semi-permanent broadband and accelerometric network in the vicinity of the dam, recording local and regional seismicity until the latter half of 2009. During this period, the recorded local seismicity comprised events with maximum magnitudes reported by the National Geographical Institute of Spain (IGN) of mbLg=2.7.

## 2 METHODS AND DATA

Dynamic deformations refer to the time-dependent strains and rotations produced in a given site due to the propagation of seismic waves. We apply and compare two different methodologies to analyze the dynamic deformations due to the local seismicity near the studied reservoir. The first one based on the Seismo-Geodetic method requiring data from a seismic array with at least three, three-component seismic stations. The second method based on a Single Station analysis and thus requiring data from only one three-component station. In the following subsections, these two methods and the input seismic data for this study are described in detail.

### 2.1 Seismo-Geodetic method

The term "Seismo-Geodetic" was first introduced by Bodin et al. (1997), to describe an imported method from geodesy to study the dynamic deformations given the displacements from seismic data. This method, introduced to calculate the deformation field at the surface, is characterized by the displacement gradient tensor obtained from a seismometric micro-array with at least three, three-component stations. It is based on the procedure described by Spudich et al. (1995) and Bodin et al. (1997), where the displacement gradient tensor $G^g_{i,j}(t)=\partial u_i(t,\boldsymbol{x})/\partial x_j \approx \Delta u_i(t,\boldsymbol{x})/\Delta x_j$ is calculated at each time from the recorded ground displacements $u_i$, i=1,2,3; j=1,2,3, and where $u_1$, $u_2$, $u_3$ are the displacements in the $x_1$=x=east, $x_2$=y=north, $x_3$=z=up respectively; the superscript "$g$" indicates the terms for the Seismo-Geodetic method. In the following, the time dependence is understood. $G^g_{i,j}$ then, is obtained by solving for each time the set of equations,

$$\mathbf{d}^m = G^g_{i,j}\mathbf{R}^m \qquad (1)$$

where $\mathbf{d}^m = \mathbf{u}^m - \mathbf{u}^0 = (u_1^m - u_1^0, u_2^m - u_2^0, u_3^m - u_3^0)$ is the relative recorded displacement $\mathbf{u}$ between the reference station (superscript "$0$") and station "$m$", and $\mathbf{R}^m = \mathbf{r}^m - \mathbf{r}^0 = (x_1^m - x_1^0, x_2^m - x_2^0, x_3^m - x_3^0)$ is the difference in spatial coordinates between these stations. At the surface, due to the stress free boundary conditions, three components of $G^g_{i,j}$ are not independent: $\partial u_1/\partial x_3 = -\partial u_3/\partial x_1$ , $\partial u_2/\partial x_3 = -\partial u_3/\partial x_2$ and $\partial u_3/\partial x_3 = \eta\ (\partial u_2/\partial x_2 + \partial u_1/\partial x_1)$   where $\eta = -\lambda/(\lambda+2\mu)$; $\lambda$ and $\mu$ being the Lamé Parameters. The system in (1) can be rewritten by setting the unknown displacements $G^g_{i,j}$ into a column vector $\boldsymbol{\gamma} = (\ \partial u_1/\partial x_1\quad \partial u_1/\partial x_2\quad \partial u_1/\partial x_3\quad \partial u_2/\partial x_1\quad \partial u_2/\partial x_2\quad \partial u_2/\partial x_3)^T$, having

$$\mathbf{d}^m = \mathbf{P}^m\boldsymbol{\gamma} \qquad (2)$$

where $\mathbf{P}^m$ is a 6X3 matrix containing the reordered differences $\mathbf{R}^m$ (e.g. Bodin et al, 1997). Then, the linear system to be solved is

$$\mathbf{d} = \begin{pmatrix} \mathbf{P}^1 \\ \mathbf{P}^2 \\ \vdots \\ \mathbf{P}^m \end{pmatrix} \boldsymbol{\gamma} \qquad (3)$$

where $m \geq 2$. This overdetermined system of equations is solved by a least squares procedure (e.g. Menke 1989; Spudich et al. 1995; Bodin et al. 1997). Once obtained the components of $G^g_{i,j}$, these are used to derive the uniform strains

$$\varepsilon_{i,j} = \frac{1}{2}\left(\frac{\partial u_i}{\partial x_j} + \frac{\partial u_j}{\partial x_i}\right) \quad (4)$$

and the rigid body rotations

$$\omega_{i,j} = \frac{1}{2}\left(\frac{\partial u_i}{\partial x_j} - \frac{\partial u_j}{\partial x_i}\right) \quad (5)$$

where again, $i$=1,2,3; $j$=1,2,3 and $u_1$, $u_2$, $u_3$ are the displacements in the x, y and z directions at a given time. Due to the stress free boundary conditions, the strain terms $\varepsilon_{1,3}=\varepsilon_{3,1}=\varepsilon_{2,3}=\varepsilon_{3,2}=0$. Positive rotation is defined in counter-clockwise sense, viewed from the positive axis.

## 2.2 Single Station method

In this method, dynamic deformations are obtained at a single site by using a set of three component velocity seismograms under the assumption of body wave incidence through the recording site. The 3D displacement gradient tensor components at surface are obtained by means of the method described in Gomberg (1997) and Singh et al. (1997). In this method, the displacement gradients used to calculate the dynamic deformations are obtained assuming the incidence of body waves through the recording site; thus, considering S waves incidence, the particle motion can be written as

$$u_{SH} = A_{SH}e^{i(\omega t - k_h \cdot r + \theta_{SH})} \;\; ; \;\; u_{SVh} = A_{SVh}e^{i(\omega t - k_h \cdot r + \theta_{SVh})} \text{ and } u_{SVv} = A_{SVv}e^{i(\omega t - k_h \cdot r + \theta_{SVv})} \qquad (6)$$

where $u_{SH}$ is the SH motion in the transverse direction, and $u_{SVh}$ and $u_{SVv}$ are the SV motions in the radial and vertical directions respectively. $A_{SH}$, $A_{SVh}$ and $A_{SVv}$ are the amplitudes in the transverse, radial and vertical directions; $\theta_{SH}$, $\theta_{SVh}$, $\theta_{SVv}$ are the phases of the incident wavefield at the surface in the measurement point; $\omega$=$2\pi f$ is the angular frequency and $i = \sqrt{-1}$. In this case, $k_h \bullet r = k_x x + k_y y$, where the magnitude of the horizontal wavenumber is $k_h$=$2\pi/\lambda_h$ and $\lambda_h$ is the horizontal wavelength. This last one can be expressed as $\lambda_h$=$TV_s/sin(\psi)$=$TV_h$, where the period is $T$=$1/f$, $V_s$ is the S wave velocity, $V_h$ is the horizontal apparent velocity and $\psi$ is the angle of wave incidence with respect to the vertical.

The complete S wave motion, propagating along the two horizontal directions (*x, y*) at surface, can be resolved in the three Cartesian directions as (e.g. Gomberg 1997)

$$u_j^x = A_j^x e^{i\theta_s} e^{i(\omega t - k_x x)} \text{ and } u_j^y = A_j^y e^{i\theta_s} e^{i(\omega t - k_y y)} \text{ ; } j=1,2,3 \qquad (7)$$

where subscripts *j=1,2,3* indicates the particle motion component of S waves in the **x** , **y** and **z** directions respectively.

The spatial derivatives of equations (*7*) are then taken in order to obtain the displacement gradients. From this, the spatial gradients $G_{i,j}^s$ (superscript "*s*" for single station) of equations (7), at the measuring site can be written as:

$$G_{xx}^s = \frac{\partial u_1}{\partial x_1} = -\frac{\partial u_1}{\partial t}\frac{\sin\phi}{V_h} = -\frac{\partial u_1}{\partial t}\frac{\sin\phi\sin\psi}{V_s} \text{ ; } G_{yx}^s = \frac{\partial u_2}{\partial x_1} = -\frac{\partial u_2}{\partial t}\frac{\sin\phi}{V_h} = -\frac{\partial u_2}{\partial t}\frac{\sin\phi\sin\psi}{V_s}$$

$$G_{zx}^s = \frac{\partial u_3}{\partial x_1} = -\frac{\partial u_3}{\partial t}\frac{\sin\phi}{V_h} = -\frac{\partial u_3}{\partial t}\frac{\sin\phi\sin\psi}{V_s} \text{ ; } G_{xy}^s = \frac{\partial u_1}{\partial x_2} = -\frac{\partial u_1}{\partial t}\frac{\cos\phi}{V_h} = -\frac{\partial u_1}{\partial t}\frac{\cos\phi\sin\psi}{V_s} \qquad (8)$$

$$G_{yy}^s = \frac{\partial u_2}{\partial x_2} = -\frac{\partial u_2}{\partial t}\frac{\cos\phi}{V_h} = -\frac{\partial u_2}{\partial t}\frac{\cos\phi\sin\psi}{V_s} \text{ ; } G_{zy}^s = \frac{\partial u_3}{\partial x_2} = -\frac{\partial u_3}{\partial t}\frac{\cos\phi}{V_h} = -\frac{\partial u_3}{\partial t}\frac{\cos\phi\sin\psi}{V_s}$$

where $\phi$ is the angle of azimuth of the incidence wavefield. Given that these expressions have an inverse dependence with Vs, as the incidence angle approaches to the vertical the spatial gradients become more sensitive to Vs. As in the previous section, due to the stress free boundary conditions, three components of $G_{i,j}^s$ are not independent at surface: $\partial u_1/\partial x_3 = -\partial u_3/\partial x_1$ , $\partial u_2/\partial x_3 = -\partial u_3/\partial x_2$ and $\partial u_3/\partial x_3 = \eta (\partial u_2/\partial x_2 + \partial u_1/\partial x_1)$ and $\eta=-\lambda/(\lambda+2\mu)$; $\lambda$ and $\mu$ are the Lamé parameters. Once obtained the components of $G_{i,j}^s$ at a given time, the uniform strains can be obtained by the equation (4) and the rigid body rotations by the equation (5).

## 2.3 Data and processing

The seismic data used in this work are the recordings from local earthquakes obtained by means of a semi-permanent network installed in 2008 by the University of Almería (UAL), using broadband and accelerometric stations. Seismic sensors were located on the surface and at underground sites in the close vicinity of the dam (Figure 2). The seismographic network installed by the UAL in the region, consisted of five Etna-Kinemetrics accelerometers and five Guralp CMG-3ESP velocity broadband seismographic stations.

Among these ten stations, an array of two broadband sensors (PIAL and PGAL) and one accelerometer (PDAL) was installed in the close proximity of the dam (Figure 2). PDAL and PIAL were installed at surface, and PGAL was placed in an underground gallery on the left-margin of the dam structure.

Given that the array consisted of two different types of seismic instruments, we performed a seismometric cluster-test to assure that the ground motions recorded at a given site were equivalent in amplitude and phase in both types of instruments, and could be jointly employed for the analysis. The details of the test procedure are explained in Appendix A. Figure 8 shows the time series displacements produced by a local event on both instruments at a single site, exhibiting a sharp agreement between them. Indeed, an overlaid plot of the time series shows them almost indistinguishable. Because of this, in this Figure we artificially shifted-up the base-line of the integrated accelerograms for a clearer comparison. This test was also done for two other earthquakes obtaining the same agreement between waveforms. Test results confirm the feasibility of using these two types of instruments for the analysis.

During the operating period of the network at the dam site (October 2008-September 2009) different local and regional earthquakes were recorded. From these, a limited number of earthquakes with magnitudes M>1.0 took place in the vicinity of the dam. Based on the assumptions for dynamic deformations using the single station method, we selected amid the shallow local earthquakes (Z<12.0 km) those that were recorded in the three stations with epicenters closer than 10.0 km from the recording sites. This selection was done in order to primarily have a vertical incidence of wave-fields at the recording station, reducing in this way the incidence of surface waves energy at the recording stations, and assuring that most of the incoming seismic energy is contained in the S wave-field. In total, 12 local earthquakes satisfied these conditions with reported magnitudes by the Instituto Geográfico Nacional (IGN) ranging from mbLg=0.8 to mbLg=2.2. The selected events occurred between November 2008 and April 2009 (Table 1). The epicentral location was first refined by integrating the data from the public catalogs of IGN and the body wave arrival times picked from the UAL network, using a modified crustal velocity structure for the region obtained by Ruiz et al. (2006a). We then performed a relocation of hypocenters by mean of the double-difference methodology proposed by Waldhauser and Ellsworth (2000). The parameters of these earthquakes are shown in Table 1.

To obtain the ground displacements for the analysis, time series were integrated in all cases in time domain. Seismograms from the broadband seismometer were time-integrated once, and time series from the accelerograph were time-integrated twice. Integrations were performed following the procedure by Iwan et al, (1985) and Boore and Bommer (2005). Time series were then band-pass filtered with a 4-pole Butterworth filter in the frequency band of 0.5 Hz<$f_c$<1.85 Hz. The 1.85 Hz upper bound of the filter was selected in order to avoid the possible spatial aliasing from signals with wavelengths significantly smaller than our array. Specifically, to obtain array gradient estimates accurate to approximately 90% of true gradients, the array dimensions must be less than approximately one quarter-wavelength of the dominant energy in the wave train (Bodin et al. 1997). Given this, the upper bound of the filter should satisfy the relation L≤$\lambda$/4=$V_s$/4$f$, where L= maximum vertical distance among stations. Satisfying this ensures that the deformation field is uniform within the array at any time. The complete dataset of earthquake recordings and filtered displacements for all earthquakes, stations and components are shown in appendix B (Online Resource 1).

Due to their relative spatial locations, the three recording stations used for the analysis are situated along an almost vertical plane which in a horizontal map projection, stations appear almost aligned. Given this for our analysis, the relative spatial locations among the recording stations with respect to the hypocentral locations of the studied earthquakes should characterize an actual three-dimensional setting. In this analysis, none of the hypocentral locations of the studied earthquakes are situated along this hypothetical plane.

## 2.4 Shallow velocity structure

Taking advantage of the spatial setting of the recording stations, and being the studied earthquakes located near the vertical projection from stations, we performed a body wave velocity analysis of the shallow layer near the dam.

Assuming vertical incidence of body waves, we calculated the mean shallow velocity structure at the dam site. PGAL station was located at a depth of $Z_{PGAL}$=203.0 m with respect to PIAL (at surface), in the left margin of the dam. We used the time picking arrivals of the P and S wavefronts at both stations, for each of the studied earthquakes, to obtain an average estimation of the S wave travel time between both stations. *Vs* velocity values were first obtained from direct S-wave arrival measurements. Given that some recordings presented long P wave-trains, we compared these velocities with those obtained from P wave arrivals assuming a Poisson ratio of *v*=0.25. The *Vs* values obtained from both measurements are summarized in Table 2. From here, the mean S wave velocity for the shallowest 203.0 m resulted as *Vs*=1.5±0.2 km/s. In an independent study in the same site based on Rayleigh-wave dispersion curve inversion, Rivas et al. (2012) assumed a 30.0 m soft layer overlaying a halfspace, and found an S-wave velocity of 0.902 km/s for the upper layer and 1.736 km/s for the bedrock. From these results, the travel-time equivalent S-wave velocity for the shallowest 203.0m would be $Vs_R$=1.527 km/s, which is consistent with our results. Hence, we assume in this study an S wave velocity of *Vs*=1.5 km/s for the upper structure. For our analysis this shallow low-velocity layer was added to the regional structure used by Ruiz et al (2008a).

## 3 RESULTS AND DISCUSSION

The dynamic displacement-gradients, strains and rotations from the Seismo-Geodetic method were computed for each studied earthquake using the seismic data from the micro-array close to the dam. The computation of all tensor components was performed considering each station (PIAL, PDAL and PGAL) as the reference site (*"0"),* having in this way three sets of dynamic tensor components. This analysis ought to be done in this way because, while relative amplitudes obtained from this method are equal no matter which station is used as reference, the phase information changes depending on the selection of the reference site.

For the Single-Station estimates, the set of wave incident angles at surface ($\phi_n$, $\psi_n$) were computed for each $n^{th}$ earthquake shown in Table 1, including the shallow velocity layer in the crustal structure, and taking into account the relative location of each $n^{th}$ hypocenter with respect to each recording site (PIAL, PDAL and PGAL). For each time and earthquake, all terms of the displacement gradient tensor were computed using equations (8) and the seismic data from each reference station. Then, the four non-vanishing terms of the computed strain tensor and the three rotational components at surface were calculated using equations (4) and (5).

Figures 3 and 4, show the comparison of time series between both methods for the six tensor components of dynamic displacement gradients at all the three stations (PIAL, PDAL and PGAL), for two selected earthquakes (891920 and 907842). Figures 5 and 6, show the comparison of the time series between both methods for the four non-vanishing terms of the strain tensor ($\varepsilon_{xx}$, $\varepsilon_{yy}$, $\varepsilon_{zz}$, $\varepsilon_{xy}$) and the three terms of the rigid body rotations ($\omega_x$, $\omega_y$, $\omega_z$), at all the three stations and same selected events. In these Figures, the results obtained by the Seismo-Geodetic method are shown with solid lines and those obtained by the single-station method are shown with dashed lines. The complete dataset of dynamic deformations results for each earthquake, station and tensor component are shown in appendix C (Online Resource 1).

From these figures it can be observed that the time histories of displacement gradients show a good agreement in amplitude and phase along the main S-wave arrival interval. In the same way, dynamic strains and rotations also show a good agreement between both methods along this same interval. Peak (absolute maximum) strain amplitudes in both cases globally varies between $10^{-9}$ and $10^{-8}$ *strain*, except $\varepsilon_{zz}$ which varies between $10^{-10}$ and $10^{-9}$, and the peak rotations between $10^{-8}$ and $10^{-7}$ *radians* for this range of magnitudes.

Results show that the maximum amplitudes on the displacement gradients are mainly observed on $\partial u_1/\partial x_2$. For this reason, the horizontal components of uniform strains from both methods systematically present larger amplitudes than the vertical ones. Rotations, in the same way, present larger amplitudes around the vertical axis, as they are given by the horizontal amplitudes of the displacement gradients. This observation makes sense due to the small incidence angles at the recording stations. As S waves distort the media in the transverse direction of propagation, the largest amplitudes of ground motions, dynamic displacement-gradients and dynamic strains tend to appear on the horizontal components.

In order to quantify the misfit between the time histories of dynamic strains and rotations from both methods, we calculated for each pair of waveforms, a normalized L2-norm of the form,

$$N_{i,j} = \frac{1}{A_{max} \sqrt{n}} \sqrt{\sum_{k=1}^{n} [\Psi_{i,j}^s(k\Delta t) - \Psi_{i,j}^g(k\Delta t)]^2} \qquad (9)$$

where $\Psi_{i,j}$ could be either $\varepsilon_{i,j}$ or $\omega_{i,j}$ ; $A_{max}$ is the absolute value of the maximum amplitude of each pair of the compared time histories, $\Delta t$ is the time increment and $n$ is the number of time samples. The norm was computed only for the approximate interval of S-wave arrivals, excluding in this way the coda from the misfit computation. Given the range of magnitudes of the studied earthquakes, we chose an interval of 2.0s after the first S-wave arrival for all earthquakes. Table 3 shows the values of $N_{i,j}$ for each term of the strain tensor and rotational component, for the three stations and 12 events studied. The strain misfits range from 1.4% in the $\varepsilon_{xx}$ component of the 885494 earthquake being PIAL as reference station, to 34.9% in the $\varepsilon_{yy}$ component of the 885470 earthquake being PGAL as reference station. On the other hand, rotational misfits range from 2.5% on the $\omega_z$ torsion component (rotation around the vertical axis; $\omega_{i,j}$ for $i{\neq}j{\neq}3$) of the 885470 earthquake for station PIAL, to 36.0% on the $\omega_x$ tilt component (rotation around the EW axis; $\omega_{i,j}$ for $i{\neq}j{\neq}1$), for earthquake 896491 of station PGAL.

The best adjustments between time histories from both methods occur during the S-wave arrival interval, which is expected from the single-station model assumptions. During this interval, because of the low incidence angles and the proximity of the studied earthquakes, ground motions are usually less altered by the scattering effects from geological irregularities. The horizontal components of strain and rotations show in general lower misfits between both methods than the vertical ones. This can be observed from Table 3, where strain misfits vary on the horizontal components from 1.4% to 13.5%, except in the case of station PGAL where for four earthquakes, the horizontal component presents larger misfits, between 24% and 34%. This particular behavior occurs only for earthquakes 885470, 885487, 885493 and 890840. As this station is located in one of the galleries of the dam, and these earthquakes occur at shallower depths, this might be due to some local velocity anomaly or a strong geological lateral variation. In the same sense, rotational misfits are in most cases below 10% on the $\omega_z$ torsion component and above this value on the tilt components. This may be partially explained because the P-wave scattered energy is usually more present on the vertical motion components, which is not explicitly considered in this single-station model. Gomberg and Agnew (1996) compared for three California earthquakes, the dynamic deformations observed by the strainmeter system at Piñon Flat, California, with seismic estimates, assuming surface wave incidence to the seismic stations at the same recording site. They obtained amplitude misfits of ±20% and phase misfits of ±10% between the observed and the seismic-estimated strains. These results show that estimates like the obtained here, adequately reproduce the order of magnitude of strains and rotations for this type of data. Our misfits show that the differences between the two methods studied here are within the expected percentages, as also of the same order of magnitude with respect to measurements. Given this and while both methods studied here have shown to be adequate for dynamic deformation evaluations, the Single-Station method might be a good option for the analysis of local seismicity in areas with a limited number of three-component stations available.

## 3.1 Scaling of peak ground dynamic deformations

Amplitudes of displacement gradients obtained in this work vary between $10^{-9}$ to $10^{-7}$

*strain*. These low amplitudes are mostly related with the low-magnitude of the earthquakes analyzed (M<2.5). The largest earthquakes reported for the Itoiz region have magnitudes between 4.8 and 5.5, generating intensities between V and IV in the Pamplona region (Ruiz et al. 2006a). Based on the probabilistic seismic hazard assessment performed for the Itoiz dam region by Rivas-Medina et al (2012), using earthquake databases for the region including events occurring in different active fault systems near the Itoiz dam, an Mw5.5 earthquake in the region near the dam can be expected.

In order to have an estimation of the possible peak dynamic deformations produced by local earthquakes up to the maximum observed magnitudes at this site (M=5.5), we performed a scaling analysis for the peak ground dynamic deformations.

Equations (8) show that ground displacement gradients ($G_{i,j}$) hold a direct relationship with the surface ground velocities $\left( \dfrac{\partial u_i}{\partial t} \right)$. Several studies have shown that in the far field, peak ground velocities ($v_{max}$) follow a general scaling relation of the form (e.g. Campbell 1997; Douglas 2002; Dost et al. 2004):

$$\log(v_{\max}) = \beta_1 + \beta_2 M + \beta_3 \log(R) + \beta_4 \log(v_{\max}^*) \qquad (10)$$

where $R$ is the earthquake-station distance (hypocentral or epicentral), $M$ is the earthquake magnitude ($M_l$, $M_s$ or $M_w$), $v^*_{max}$ is the near field peak ground velocity, and $\beta_1$, $\beta_2$, $\beta_3$, $\beta_4$ are constants to be evaluated for specific sites, ranges of magnitudes and epicentral distances. For a fixed spatial relationship ($R$=const) between a given earthquake and a given recording station, the parameters $\psi$, $\phi$ and the velocity $Vs$ in (8) become constants, and $G_{i,j}$ acquire a linear relationship with magnitude:

$$\log\{(G_{i,j})_{\max}\} = \gamma_{1,i,j} + \gamma_{2,i,j} M + \gamma_{3,i,j} \log(G_{i,j})_{\max}^* \qquad (11)$$

where $M$ is again the earthquake magnitude ($M_l$, $M_s$ or $M_w$), $G^*_{i,j}$ is a near field peak ground dynamic deformations term, and $\gamma_{1,i,j}$, $\gamma_{2,i,j}$, $\gamma_{3,i,j}$ are constants to be evaluated; here sum convention does not apply.

The studied earthquakes belong approximately to the same source volume at similar distances. Given this, the soil structure and the spatial setting between the source and the reference station can be assumed fixed and equation (11) applies to $G_{i,j}$. Constants $\gamma_{1,i,j}$, $\gamma_{2,i,j}$, $\gamma_{3,i,j}$ may be evaluated by regression analysis when having enough observed data. Due to the limited number of earthquakes in our analysis, we performed a synthetic modelling of surface ground dynamic deformations (uniform strains and rigid-body rotations) fitting the observed data, in order to estimate the intervals of $(G_{i,j})_{max}$ values for different magnitudes, up to the maximum expected magnitude in the region ($M_w$=5.5). To do this, we assumed for the crust a half-space with a Poisson's ratio of $v$=0.25. Synthetic displacements were computed by the Discrete Wavenumber method (Bouchon 1979), and the spatial derivatives for displacement gradients and the computation for dynamic deformations were obtained by a first-order finite difference approach, following the methodology by Santoyo (2014).

Figures 7a-7d shows the peak dynamic uniform strains and Figures 7e-7g shows the peak rigid-body rotations for 10,015 simulated earthquakes, for a magnitude interval between Mw=0.5 and Mw=5.5. Here we assume as the source hypocenter the centroid of the studied earthquakes and each station as the receiver site. Observed peak dynamic deformations from the 12 studied earthquakes are also shown by black stars in circles. For this case, the mbLg reported magnitudes were converted to $M_w$ by the relation for the Iberian Peninsula $M_w$=0.311+0.637$mbLg$+0.061$mbLg^2$ obtained by Rueda and Mezcua (2002).

As it can be observed, the simulated peak dynamic deformations follows expression (11) up to an $M_w$≈3.0. Above this value, given the short source-receiver distance assumed in our simulations ($R$=4.1 km), the slope slightly changes due to the influence of the near field terms.

Figures 7a-7g show that synthetic results fit reasonably well with the observed values for the magnitude range of the observed earthquakes. A possible overestimation of deformations might result in the $\varepsilon_{zz}$ component (Figure 7g). This may be due to the crustal structure assumed as this component is computed based on the terms ($\varepsilon_{xx}$, $\varepsilon_{yy}$)

and the elastic constants at surface. Uniform strain amplitudes for the magnitude interval $M_w[0.8,2.2]$ varies between $10^{-9}$ and $10^{-8}$ *strain*, and rigid body rotations between $10^{-9}$ and $10^{-7}$ *radians*. From these results, a local $M_w$=5.5 magnitude earthquake might then produce values for peak uniform strains between $10^{-6}$ and $10^{-5}$ *strain* and $10^{-5}$ and $10^{-3}$ *radians* at the Itoiz dam site. Santoyo (2014) obtained for the Lorca 2011 ($M_w$5.2) earthquake, peak dynamic strain amplitudes of $10^{-5}$ *strain* and peak dynamic rotations of $10^{-3}$ *radians*, for similar hypocentral distances. Those results support the order of magnitude of strains and rotations obtained in our estimates.

Soil failures related with dynamic deformations may occur at a great diversity of strain and rotation amplitudes, depending on the type of subsoil and saturation conditions. As reference values, nonlinear behaviour in unsaturated soils due to seismically induced dynamic deformations has been observed for strain values of $10^{-5}$, with liquefaction for strains of $10^{-4}$ (e.g. Drnevich and Richart 1970; Dobry et al. 1981; Tokimatsu and Seed 1987). Laboratory tests on San Francisco Bay muds, which failed and liquefied during the 1989 Loma Prieta earthquake (Mw6.9), show loss of rigidity for cyclic strains of $10^{-5}$ (Stewart and Hussein 1992; Day 2002). However, Singh et al. (1997) noted that the soft-clays of the shallow sediments in the Mexico City Valley, behave almost linearly at shear strains as high as 1%. They also found that during the 19 September, 1985 earthquake, failures of water lines pipes may have been produced at strains of $10^{-3}$. In the case of large structures, possible failures from seismic loads in dams depends on the dam structure type and geometry, the induced hidrodynamic loads and the foundation behaviour, among others (e.g. Chopra et al. 1980; Hall 1988; Bouaanani et al. 2004). Because of this, explicit values for strain amplitudes leading to possible failures in this particular gravity dam should be specifically studied.

The source and propagation-path characteristics of seismicity near reservoirs (possible induced seismicity and relatively high intrinsic attenuation) might also play an important role on the peak dynamic deformations, as they are directly related with the peak ground velocities. High peak values on the ground motions may produce high amplitudes on the ground dynamic deformations; however, low magnitude earthquakes and strong intrinsic attenuation near water reservoirs, might be masking the actual deformations amplitudes. This might be observed in Figure 7 where observed peak deformations remain below the maximum synthetic expected values. A more detailed analysis on the effects produced by both types of seismicity on the peak and spectral values should be of interest for future works on this subject.

# 4 CONCLUSIONS

We computed the near-source surface 3D displacement-gradients and dynamic deformations from the analysis of the ground motion time histories due to the local seismicity near the Itoiz reservoir. The dynamic deformation field was calculated by two different methodologies: the Seismo-Geodetic method and the Single-Station method for the displacement gradients. Results from both methods were compared and analyzed getting misfits for uniform strains between 1.4% and 35.0% and misfits for rigid rotations from 2.5% to 36.0%. The shallow 1D velocity structure was estimated from the seismic data by modeling the body P and S wave travel times resulting in an S wave surface average velocity of 1.5km/s. Observed amplitudes of displacement gradients vary in the range of $10^{-8}$ to $10^{-7}$ strains, peak uniform strain amplitudes between $10^{-10}$ and $10^{-8}$ *strain*, and peak rigid rotations between $10^{-8}$ and $10^{-7}$ *radians*.

Based of these results, a synthetic numerical modeling was performed to estimate the peak dynamic deformations for the largest magnitude observed within this region (Mw=5.5). For this magnitude, synthetic computations produce values for peak uniform strains between $10^{-6}$ and $10^{-5}$ *strain* and $10^{-5}$ and $10^{-3}$ *radians*. Soil and lifeline failures related with dynamic deformations may occur at a great diversity of strain and rotation amplitudes, depending on the type of subsoil and saturation conditions. As reference values, nonlinear behaviour in unsaturated soils due to seismically induced dynamic deformations has been observed for strain values of $10^{-5}$, with liquefaction of soils for strains of $10^{-4}$. Our estimates for peak dynamic deformations show that these values could be reached at the Itoiz dam due to a local M5.5 earthquake.

The methods studied here have shown to be adequate to evaluate the surface dynamic deformations in these ranges of magnitudes and environment. Our results show that the Single-Station method might be a good option for the analysis of local seismicity, where few three-component stations are available. Estimates of free-field dynamic deformations from seismicity close to large human infrastructures are becoming necessary towards a better understanding of their damaging effects, as well as to evaluate its relevance for seismic hazard assessment. Here we show that the calculation of dynamic deformations from small magnitude seismicity could serve as a proxy to estimate the order of magnitude of the dynamic deformations for larger events. This study can contribute to extend the applicability of these methodologies to other sites of engineering interest.

**AKNOWLEDGMENTS**

We wish to thank Confederación Hidrográfica del Ebro (CHE) for giving access to their facilities. We thank Miguel Herraiz for its helpful comments on this study. This work was partially supported by Secretaría General para el Territorio y la Biodiversidad from Ministerio de Medio Ambiente, Rural y Marino, Spain, under grant 115/SGTB/2007/8.1, by EU with FEDER and by the research team RNM-194 of Junta de Andalucía, Spain. This work was partially done while M.A.S. was under a Ramón y Cajal grant from MICIN-Spain, and while A.G.J. was under a Juan de la Cierva grant, Spain.

**Figure captions**

**Fig 1** Location and geologic setting of the studied area. NPZ=North Pyrenean Zone; PAZ=Paleozoic Axial Zone; SPZ=South Pyrenean Zone; NPF=North Pyrenean Fault; PBM=Paleozoic Basque Massifs; IR=Itoiz Reservoir. The solid triangle indicate Pamplona city. Inset: General location of the region

**Fig 2** Location of recording stations relative to the dam wall (gray polygon). Seismic stations are shown by inverted triangles. PIAL and PDAL stations are located at surface. PGAL station location is shown by its vertical projection at the surface. Epicentral locations of the studied earthquakes are shown by gray circles. Solid Ellipses show the horizontal relocation uncertainties. UTM coordinates are shown with respect to the 3.0ºW central meridian (UTM zone 30T).

**Fig 3** Comparison of displacement gradients obtained by the two studied methods for earthquake 891920. Results from the Seismo-Geodetic method are shown with solid lines. Results obtained by Single Station estimates are shown with dashed lines. Displacement gradients are shown in *strain* units. **a.** Comparison of displacement gradients, setting station PIAL as the reference site. **b.** Same as Figure 3a setting station PDAL as the reference site. **c.** Same as Figure 3a setting station PGAL as the reference site.

**Fig 4** Comparison of displacement gradients obtained by the two studied methods for earthquake 907842. Results from the Seismo-Geodetic method are shown with solid lines. Results obtained by Single Station estimates are shown with dashed lines. Displacement gradients are shown in *strain* units. **a.** Comparison of displacement gradients, setting station PIAL as the reference site. **b.** Same as Figure 3a setting station PDAL as the reference site. **c.** Same as Figure 3a setting station PGAL as the reference site.

**Fig 5** Comparison of dynamic deformations (uniform strains and rigid body rotations) obtained by the two studied methods for earthquake 891920. Results from the Seismo-Geodetic method are shown with solid lines. Results obtained by Single Station estimates are shown with dashed lines. Uniform strains are in *strain* units and rotations in *radians*. Positive rotations are in counter clockwise direction around axes. **a.** Comparison of displacement gradients, setting station PIAL as the reference site. **b.** Same as Figure 5a setting station PDAL as the reference site. **c.** Same as Figure 5a setting station PGAL as the reference site.

**Fig 6** Comparison of dynamic deformations (uniform strains and rigid body rotations) obtained by the two studied methods for earthquake 907842. Results from the Seismo-Geodetic method are shown with solid lines. Results obtained by Single Station estimates are shown with dashed lines. Uniform strains are in *strain* units and rotations in *radians*. Positive rotations are in counter clockwise direction around axes. **a.** Comparison of displacement gradients, setting station PIAL as the reference site. **b.** Same as Figure 5a setting station PDAL as the reference site. **c.** Same as Figure 5a setting station PGAL as the reference site.

**Fig 7** Peak dynamic deformations for 10,015 synthetic earthquakes with magnitudes between Mw=0.8 and Mw=5.5 (dots). Observed peak dynamic deformations are shown with black stars within circles. See text for the source-receiver and crustal setting. 7a-7d: peak dynamic uniform strains $\varepsilon_{xx}$ , $\varepsilon_{xy}$ , $\varepsilon_{yy}$ , $\varepsilon_{zz}$ respectively.  7e-7f: peak dynamic rigid body rotations $\omega_x$ , $\omega_y$ , $\omega_z$ respectively

**Fig 8** Comparison between the displacement recordings obtained by the broadband sensor time-integrated once (solid line), and the recordings obtained by the accelerograph, time-integrated twice (dashed line). Recordings are from an Almería earthquake (mbLg=3.8) with an epicentral distance of 45.0km. Displacement amplitudes are shown in cm.

**Online Resource 1 captions**

**Appendix B**. Complete dataset of used earthquake recordings for each earthquake, station and component. Earthquake is identified by its event number in Table 1. On the upper left-hand dataset, we show the unfiltered EW, NS and Vertical components of velocities at station PIAL. In a similar way, upper centre dataset correspond to the same components at station PGAL and upper right-hand dataset correspond to station PDAL. The bottom left-hand side datasets for each earthquake correspond to the filtered displacements in EW, NS and Vertical components at station PIAL. In similar way, bottom centre dataset correspond to the same components at station PGAL, and bottom right-hand dataset correspond station PDAL. Results are plotted in ascending order of earthquake event number.

**Appendix C.** Complete dataset of dynamic deformations results. For each earthquake identified by the event number presented in Table 1, top left-hand panel shows the results obtained for all the non-vanishing dynamic deformations tensor components at station PIAL. The time series corresponding to the seismo-geodeic method are shown with solid lines and time series corresponding to the single station method are shown with dashed lines. In a similar way the right-hand panel shows results fro station PDAL and bottom panel, the results for station PGAL. Results are plotted in ascending order of earthquake event number.

**APPENDIX A**

The seismic array employed for this analysis consisted of two different types of seismic instruments: a Guralp CMG-3ESP broadband seismometer and a Triaxial EpiSensor force balance accelerometer with Etna-Kinemetrics. To assure that the ground motions recorded at a given site were equivalent in amplitude and phase in both types of instruments, we performed a seismometric cluster-test. To do this, we installed side by side, both types of sensors (the accelerograph and the broadband sensor), to record during one-month, the natural seismicity at the Seismo-Lab of the University of Almería. Then, to test the equivalence of seismic data the ground motions obtained by both instruments were compared in the displacement domain.

During the test, we recorded four local earthquakes with magnitudes between mbLg=2.7 and mbLg=3.8. For the waveform comparison, we first integrated in time domain the recorded time series in order to obtain the ground motion displacements. Integrations were performed following the procedure by Iwan et al, (1985) and Boore and Bommer (2005) which allows recovering the velocities and displacements from accelerations in a frequency band between 0.0Hz and 50.0Hz, and recovering the displacements from the broadband velocities between 0.0083Hz and 50.0Hz. The recordings obtained by the broadband sensor were time-integrated once, and the recordings obtained by the accelerograph were time-integrated twice.

After integration, due to the difference in the low-frequency response of the employed sensors (DC for accelerograph and low-corner frequency $f_c^l$=1/120 s for the broadband sensor), we applied a 3-pole Butterworth high pass filter ($f_c$=0.015 Hz) to both signals to mainly to remove the DC component of the accelerographic sensor. As the high frequency response of both sensors is approximately the same ($f_c^h$ =50 Hz), no low-pass filtering at high frequencies was applied.

Figure 8 shows the comparison between the EW, NS and vertical displacement recordings for an mbLg=3.8 event with epicentral distance of 42.0 km. As it can be observed, the comparison shows that displacements from both instruments are almost identical; actually, when recordings are plotted superimposed the time series becomes almost indistinguishable. Given this, to allow a clearer comparison between waveforms we artificially shifted-up the base-line of the integrated accelerograms in Figure 8. From here it can be observed that the seismic data recorded by both instruments are indeed equivalent in amplitude and phase assuring that data may be indistinctly used for our purpose in this frequency range. This comparison was also done for two additional earthquakes with similar hypocentral distances and magnitudes obtaining same results.

TABLE 1. PARAMETERS OF EARTHQUAKES USED IN THIS STUDY.

| Event [§] | Date | Time | M* | Lon** | Lat [#] | D [‡] |
|---|---|---|---|---|---|---|
| 885470 | 11/18/2008 | 20:29:21 | 1.8 | 1.3479 | 42.7969 | 3.3 |
| 885487 | 11/19/2008 | 2:15:24 | 1.6 | 1.3478 | 42.7959 | 2.9 |
| 885493 | 11/19/2008 | 3:03:58 | 1.2 | 1.3481 | 42.7942 | 3.4 |
| 885494 | 11/19/2008 | 5:00:37 | 0.8 | 1.3452 | 42.7929 | 3.9 |
| 890151 | 12/20/2008 | 15:33:09 | 1.2 | 1.3706 | 42.8081 | 4.4 |
| 890543 | 12/23/2008 | 18:51:15 | 1.4 | 1.3450 | 42.7955 | 4.2 |
| 890840 | 12/25/2008 | 4:08:47 | 1.6 | 1.3443 | 42.7945 | 3.7 |
| 891920 | 1/9/2009 | 2:45:15 | 1.3 | 1.3499 | 42.7969 | 4.1 |
| 891951 | 1/9/2009 | 5:04:01 | 2.0 | 1.3624 | 42.7938 | 6.9 |
| 893160 | 1/19/2009 | 3:04:37 | 1.4 | 1.3468 | 42.7956 | 3.8 |
| 896491 | 2/3/2009 | 16:24:39 | 2.2 | 1.3742 | 42.8080 | 4.1 |
| 907842 | 4/16/2009 | 21:45:58 | 1.0 | 1.3720 | 42.8071 | 4.3 |

Notes: § Earthquake number (assigned by IGN-Spain); * mbLg magnitude (IGN); **Longitude West (this work); # Latitude North (this work); ‡ Hypocentral depth in km (this work).

TABLE 2. COMPUTED S-WAVE VELOCITIES FOR THE SHALLOW STRUCTURE .

| Event | Vp | $Vs_D$‡ | $Vs_P$* |
|-------|------|------|------|
| 885470 | 2.18 | 1.61 | 1.25 |
| 885487 | 2.64 | 1.57 | 1.52 |
| 885493 | 2.51 | 1.65 | 1.45 |
| 885494 | 2.74 | 1.27 | 1.58 |
| 890151 | 2.26 | 1.56 | 1.30 |
| 890543 | 2.54 | 1.34 | 1.46 |
| 890840 | 2.94 | 1.59 | 1.70 |
| 891920 | 3.31 | 1.84 | 1.95 |
| 891951 | 2.90 | 1.56 | 1.65 |
| 893160 | 3.08 | 1.66 | 1.76 |
| 896491 | 2.54 | 1.45 | 1.46 |
| 907842 | 2.54 | 1.56 | 1.46 |

Notes: Vp= Observed P wave velocity (km/s). $Vs_D$‡ S wave velocity from direct S wave arrival (km/s); $Vs_P$*= S wave velocity obtained from Vp (Vs=Vp/√3) (km/s).

TABLE 3.  MISFIT PERCENTAGES FOR THE STRAIN TENSOR COMPONENTS

| Tensor comp | 885470 | 885487 | 885493 | 885494 | 890151 | 890543 | 890840 | 891920 | 891951 | 893160 | 896491 | 907842 |
|---|---|---|---|---|---|---|---|---|---|---|---|---|
| *PIAL station* | | | | | | | | | | | | |
| $\varepsilon_{xx}$ | 4.2 | 3.3 | 4.1 | 1.4 | 2.6 | 3.2 | 3.2 | 5.5 | 2.0 | 2.9 | 4.9 | 3.8 |
| $\varepsilon_{yy}$ | 13.2 | 4.9 | 8.7 | 12.6 | 8.5 | 2.7 | 8.0 | 4.8 | 8.6 | 6.1 | 7.9 | 5.1 |
| $\varepsilon_{zz}$ | 2.8 | 2.9 | 2.9 | 5.5 | 13.0 | 3.2 | 2.8 | 5.2 | 4.6 | 2.8 | 13.5 | 2.9 |
| $\varepsilon_{xy}$ | 5.5 | 3.6 | 2.2 | 5.3 | 2.6 | 1.5 | 2.3 | 4.3 | 5.6 | 6.2 | 6.5 | 1.5 |
| $\omega_{x}$ | 8.8 | 13.1 | 11.2 | 12.4 | 9.7 | 10.3 | 21.6 | 6.7 | 16.6 | 25.3 | 31.6 | 13.8 |
| $\omega_{y}$ | 10.7 | 24.4 | 23.8 | 10.7 | 17.0 | 11.7 | 22.3 | 9.7 | 13.5 | 15.4 | 12.2 | 14.1 |
| $\omega_{z}$ | 3.4 | 2.5 | 2.7 | 5.1 | 3.7 | 3.8 | 8.4 | 25.3 | 3.7 | 2.9 | 4.6 | 4.5 |
| *PDAL station* | | | | | | | | | | | | |
| $\varepsilon_{xx}$ | 4.3 | 6.3 | 5.6 | 4.9 | 2.3 | 3.4 | 1.5 | 5.8 | 4.3 | 15.0 | 5.8 | 4.0 |
| $\varepsilon_{yy}$ | 2.6 | 6.4 | 4.9 | 4.9 | 5.6 | 2.4 | 4.6 | 2.3 | 8.1 | 2.2 | 3.1 | 8.7 |
| $\varepsilon_{zz}$ | 3.1 | 5.1 | 4.1 | 4.0 | 12.1 | 4.3 | 3.8 | 7.9 | 3.7 | 5.8 | 9.0 | 3.6 |
| $\varepsilon_{xy}$ | 6.9 | 3.9 | 8.9 | 8.6 | 6.0 | 3.2 | 6.8 | 4.3 | 3.4 | 8.1 | 3.4 | 3.1 |
| $\omega_{x}$ | 14.0 | 16.9 | 17.4 | 16.9 | 23.3 | 9.2 | 17.6 | 13.7 | 15.1 | 33.2 | 16.1 | |
| $\omega_{y}$ | 12.4 | 16.2 | 22.6 | 12.1 | 14.4 | 26.8 | 10.3 | 9.4 | 9.3 | 9.8 | 12.2 | 17.2 |
| $\omega_{z}$ | 3.5 | 9.2 | 4.8 | 4.5 | 3.8 | 7.5 | 25.7 | 6.0 | 3.3 | 8.8 | 8.6 | 4.2 |
| *PGAL station* | | | | | | | | | | | | |
| $\varepsilon_{xx}$ | 6.2 | 5.6 | 6.2 | 5.5 | 9.8 | 3.1 | 7.9 | 9.8 | 4.2 | 3.5 | 7.0 | 3.6 |
| $\varepsilon_{yy}$ | 34.9 | 34.4 | 24.3 | 8.4 | 4.6 | 14.0 | 24.6 | 2.9 | 4.7 | 3.8 | 3.4 | 8.2 |
| $\varepsilon_{zz}$ | 5.0 | 4.4 | 4.8 | 5.5 | 3.8 | 3.5 | 3.6 | 5.2 | 7.2 | 9.6 | 9.8 | |
| $\varepsilon_{xy}$ | 7.6 | 2.4 | 2.6 | 8.9 | 2.3 | 3.5 | 3.8 | 3.6 | 5.0 | 5.5 | 12.9 | 2.7 |
| $\omega_{x}$ | 9.9 | 15.0 | 15.8 | 10.1 | 10.7 | 17.1 | 16.6 | 9.8 | 13.6 | 15.1 | 36.5 | 17.4 |
| $\omega_{y}$ | 7.0 | 15.9 | 18.4 | 8.6 | 11.3 | 16.2 | 17.4 | 18.8 | 16.6 | 15.6 | 14.8 | 17.5 |
| $\omega_{z}$ | 7.5 | 6.9 | 7.3 | 17.3 | 9.6 | 8.9 | 11.4 | 11.6 | 9.6 | 8.6 | 9.4 | 8.9 |

Notes: First row indicates the event number. *PIAL, PDAL* and *PGAL station* refer to the results obtained assuming that  given station as the reference site. $\omega_x=\omega_{i,j}$ for $i{\neq}j{\neq}1$; $\omega_y=\omega_{i,j}$ for $i{\neq}j{\neq}2$; $\omega_z=\omega_{i,j}$ for $i{\neq}j{\neq}3$.

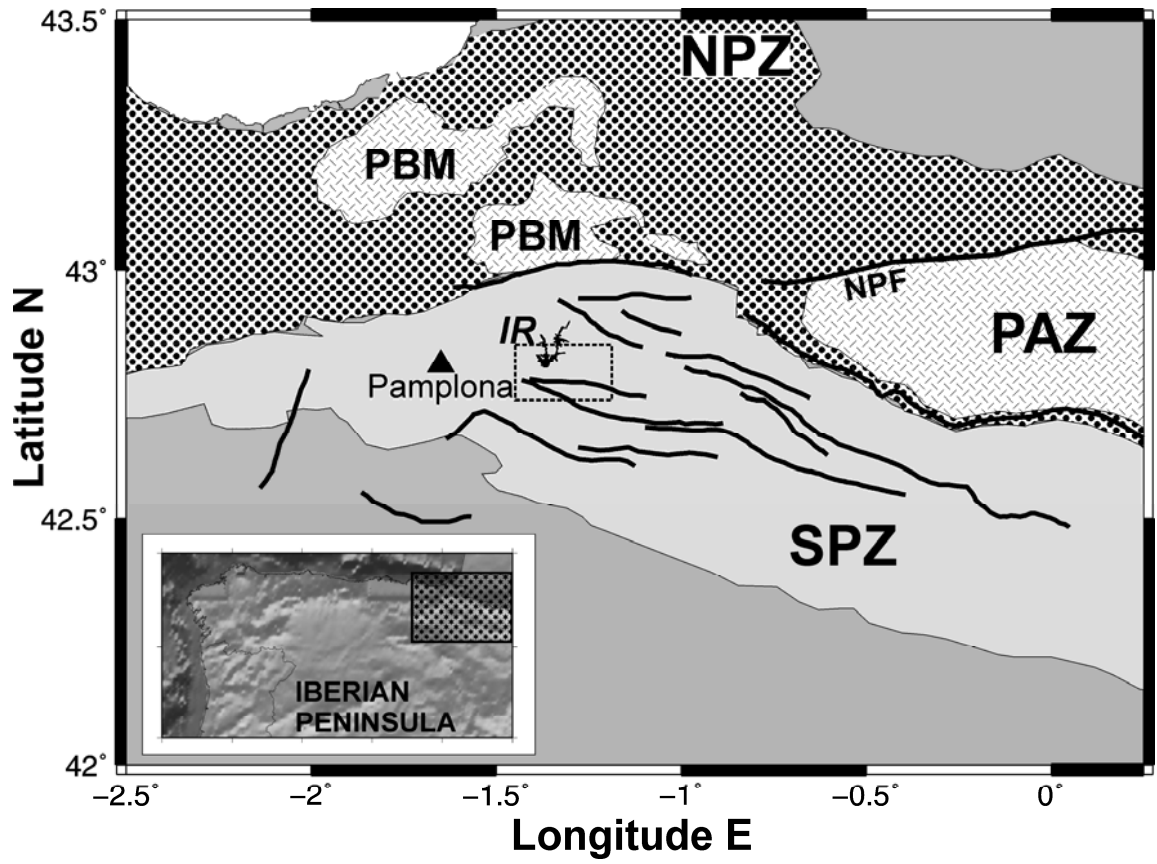

Figure 1

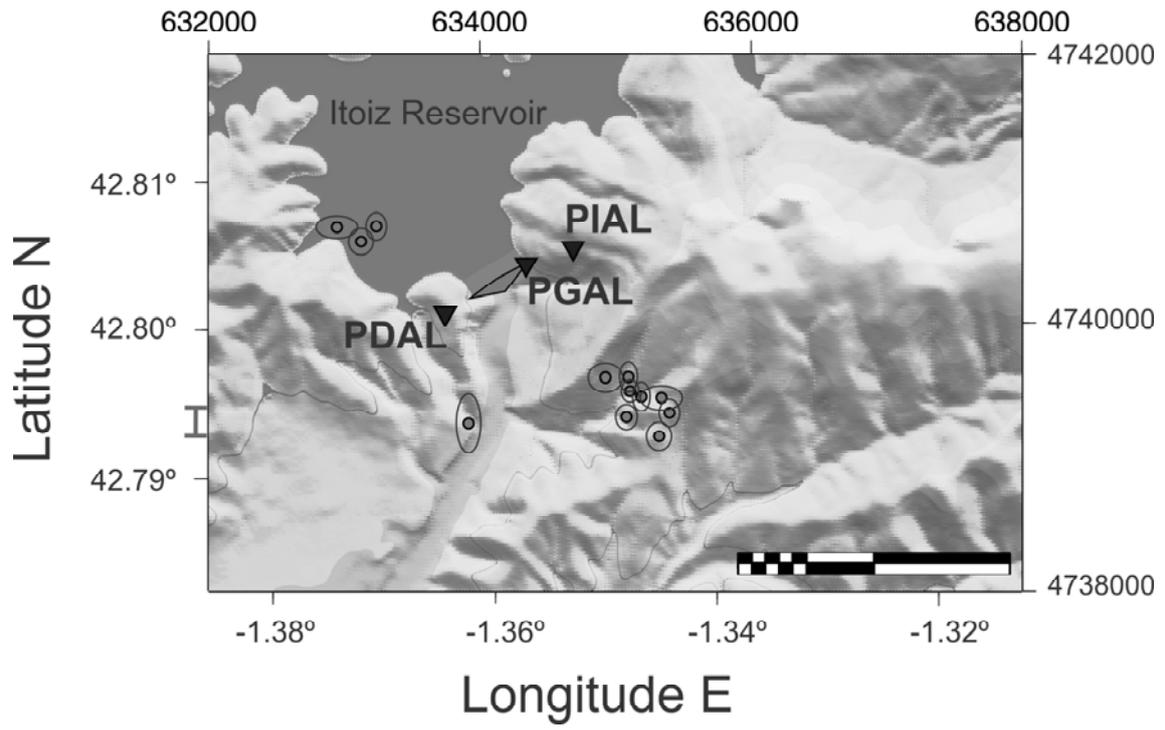

Figure 2

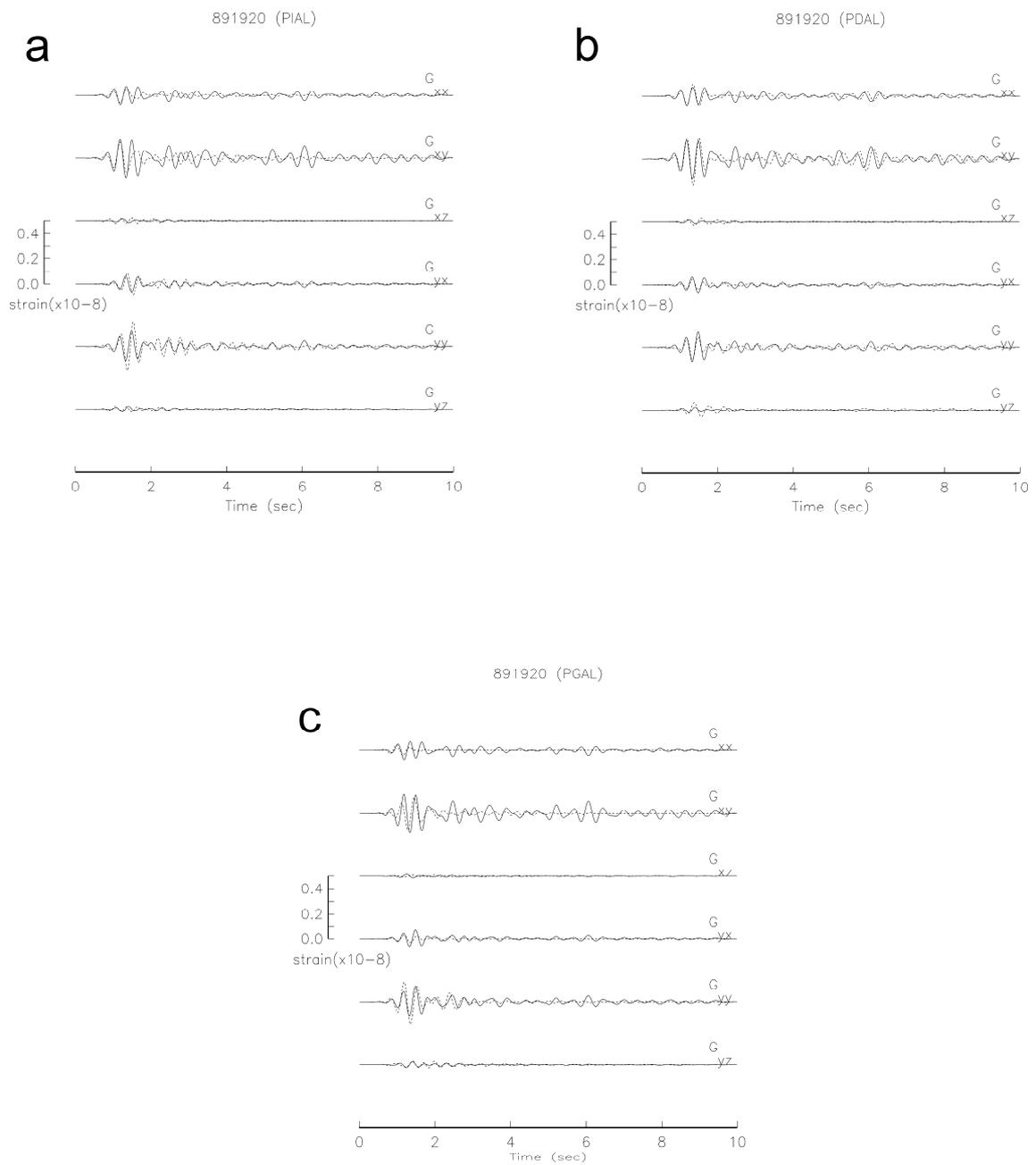

Figure 3

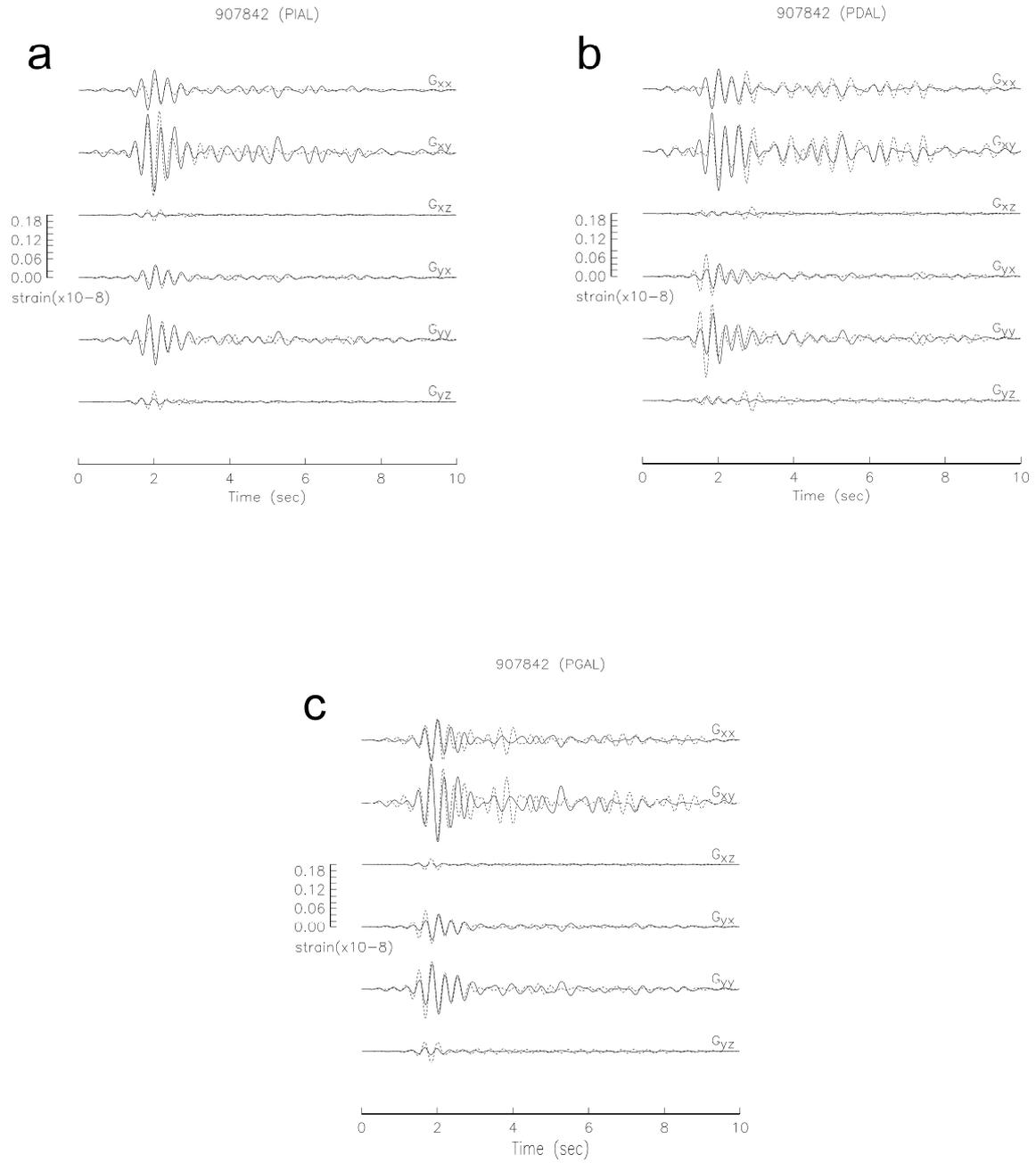

Figure 4

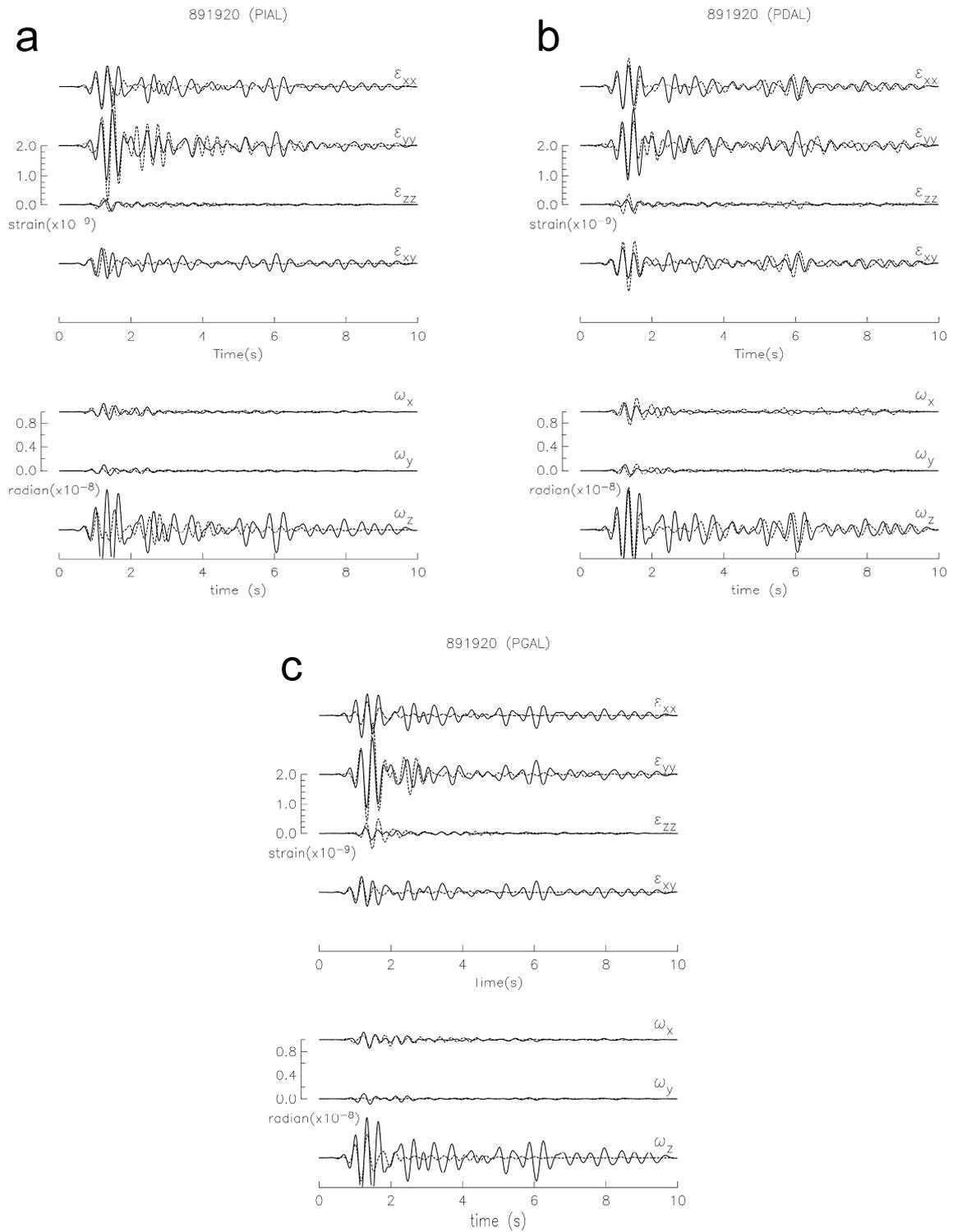

Figure 5

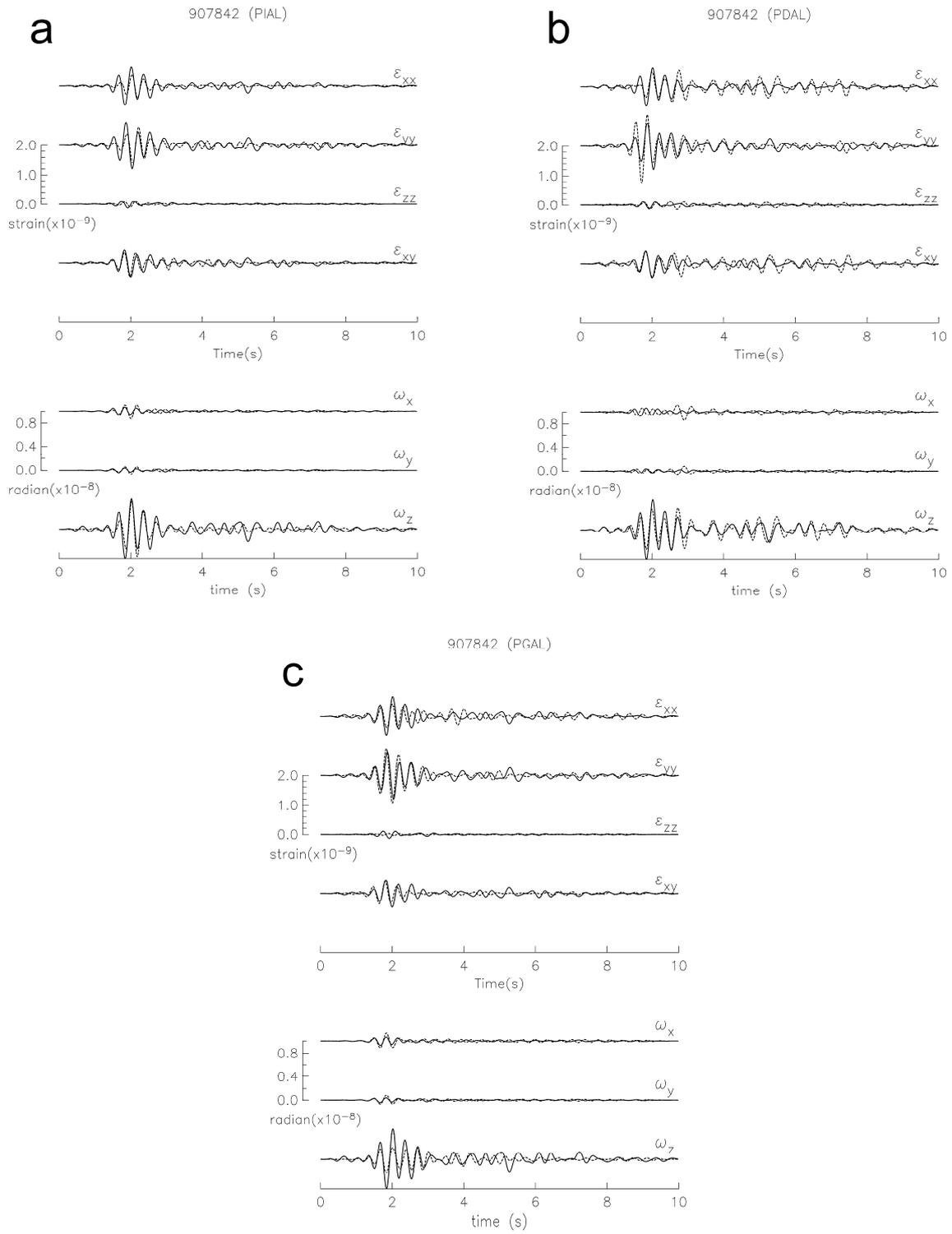

Figure 6

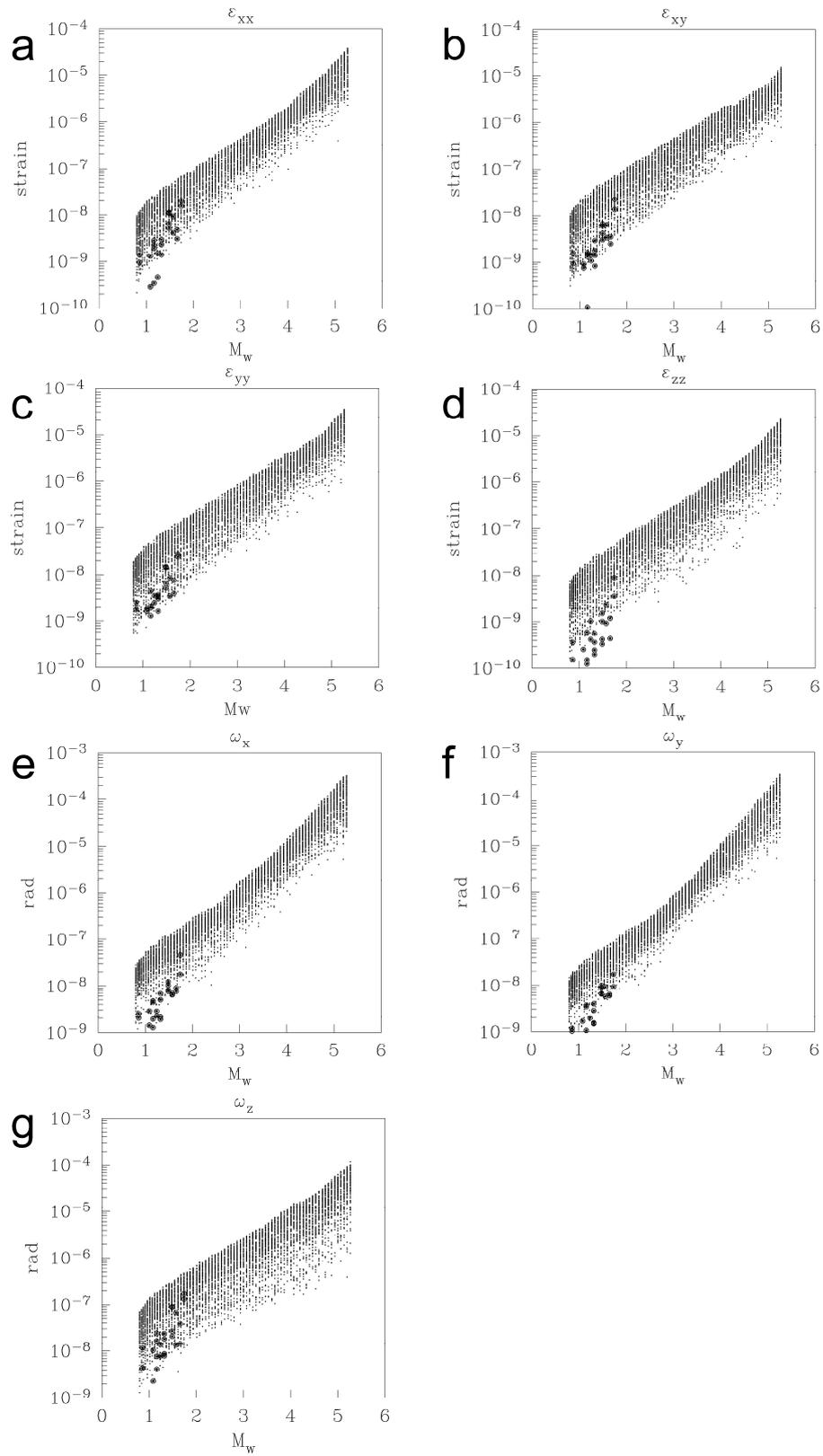

Figure 7

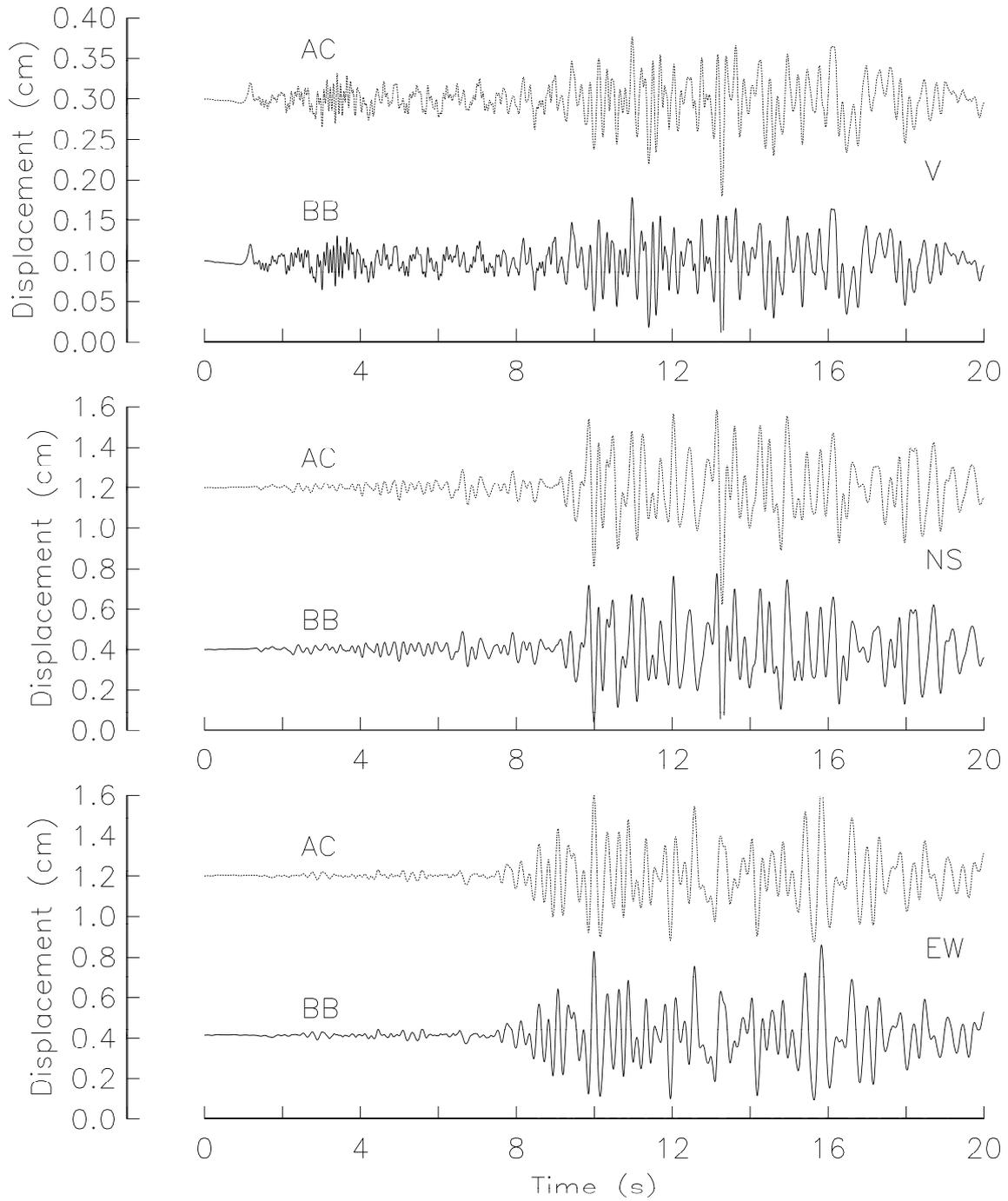

Figure 8